\documentstyle[prl,aps,epsf,multicol]{revtex}

\begin{document}
\preprint{MA/UC3M/01/1996}
\draft
\title{Fluid mixtures of parallel hard cubes}
\author{Jos\'e A.\ Cuesta\cite{email}}
\address{Departamento de Matem\'aticas, Escuela Polit\'ecnica Superior,
Universidad Carlos III de Madrid, \\ c/ Butarque, 15, 28911 --
Legan\'es, Madrid, Spain}

\maketitle

\begin{abstract}
The direct correlation function of a fluid mixture of parallel hard
cubes is obtained by using Rosenfeld's fundamental measure
approximation. This approximation is thermodynamically consistent
(compressibility and virial equations of state are equal) and predicts
a spinodal instability of the binary mixture for large-to-small side
ratio larger than roughly 10, in qualitative agreement with simulations
on the lattice version of the model. In two dimensions the
system never demix, also in agreement with the simulations.
\end{abstract}


\begin{multicols}{2}
\narrowtext

The theory of liquids is still looking for its `Ising model', i.e.\ an
exactly solvable model which provides detailed information on the
mechanisms involved in liquid behavior, and against which approximate
theories could be tested. Instead the situation is right the opposite,
and the only analytical knowledge we have about liquid models has been
obtained within the Percus-Yevick (PY) closure approximation of the
Ornstein-Zernike integral equation, and consists of
the direct correlation function (DCF) of a fluid mixture of hard spheres
with an arbitrary number of component species \cite{lebowitz}, and that
of a one-component fluid of adhesive spheres \cite{baxter}. No other
approximation has provided any new analytical result yet. Recently a
very promising approximate scheme has been proposed: the {\em fundamental
measure theory} (FMT) \cite{rosenfeld1}, which, as in the scaled
particle theory, is based upon an interpolation between the low and high
density limits of the fluid. By an appropriate choice of geometric
measures, the FMT provides the following expression for the DCF of a
fluid mixture of hard convex particles:
\begin{equation}
c_{\mu\nu}({\bf r})=[\chi_0+\chi_1R_{\mu\nu}({\bf r})+
  \chi_2S_{\mu\nu}({\bf r})+\chi_3V_{\mu\nu}({\bf r})]
  f_{\mu\nu}({\bf r}),
\label{FMT-DCF}
\end{equation}
where $V_{\mu\nu}({\bf r})$, $S_{\mu\nu}({\bf r})$ and
$R_{\mu\nu}({\bf r})$ are, respectively, the volume, surface, and mean
radius of the overlap region of particles of species $\mu$ and $\nu$
whose center-to-center vector is ${\bf r}$; $f_{\mu\nu}({\bf r})$ is the
Mayer funcion, and the $\chi_j$s
are density-dependent coefficients which can be obtained by means of the
scaled particle prescription. When applied to a mixture of hard spheres,
Eq.\ (\ref{FMT-DCF}) reproduces the PY solution; besides, the same
scheme
can be generalized to any odd dimension and thus obtain the DCF of
mixtures of hyperspheres (in one dimension it reproduces the exact DCF).
In spite of this remarkable results, the FMT has never been applied to
any molecular shape other than spheres, and it has recently been shown
that, in general, Eq.\ (\ref{FMT-DCF}) cannot be obtained for arbitrary
molecular shapes without further approximations \cite{rosenfeld2}.

While the one-component hard-sphere model has provided an excellent
starting point for the study of simple liquids, either as a reference
system for perturbation theories \cite{hansen} or as an ingredient for
density funcionals of inhomogeneous liquids \cite{evans}, the situation
is far poorer for additive fluid mixtures, due to the fact that the PY
DCF predicts stability of a binary mixture of asymmetric hard spheres
for any diameter ratio and any densities of the
components \cite{lebowitz}.
This places the demixing transition out of the scope of an analytical
treatment. Furthermore, this fact settled the belief that, contrary to
what happens in other phase transitions (such as freezing), demixing
cannot be driven by a pure entropic mechanism. This conclusion was
questioned by Biben and Hansen \cite{biben}, who found, by means of a
numeric solution of the Ornstein-Zernike equation with a more accure
closure relation, that a mixture of additive hard spheres of diameter
ratio larger than 5 and similar concentrations demixes. Further evidence
of this fact has subsequently been provided by means of alternative
approximations \cite{lekkerkerker}. The question of wether a purely
entropic mechanism can account for phase separation was affirmatively
answered by Frenkel and Louis \cite{frenkel} by providing the exact
solution of a two-dimensional lattice model of two types of particles.
For continuous three-dimensional models the controversy is still open.
Simulations cannot yet help because for too asymmetric mixtures
relaxation times are prohibitively large \cite{biben}. Nevertheless, a
recent simulation of a three-dimensional lattice model of hard
cubes \cite{dijkstra} provides strong evidence of phase separation for
mixtures of cubes of side ratio 3, while no such an evidence exists
when the side ratio is 2 (in both cases the side of the small cubes is
2 lattice spacings). Besides its computational simplicity, this model is
interesting in its own because even in the mixed phase in can completely
fill the whole space at close packing, and therefore there is no trivial
volume-driven phase separation. For the continuous version of the
model, the fourth order virial expansion of the pair correlation
function diverges at contact when the side ratio goes to
infinity \cite{dijkstra}, thus suggesting the existence of a
demixing transition.

In this letter I will show how the FMT provides an expression for the
DCF of a mixture of parallel hard cubes. By means of this function the
free-energy of the mixture can be computed, and its concavity
requirement leads to a spinodal instability for large enough side
ratio.

The starting point of the FMT is the choice of the geometric measures.
This can be easily done if the Fourier transform of the Mayer function
can be expressed as a sum of products of single particle functions. In
our case, the Mayer function of two parallel hard cubes of sides
$\sigma_{\mu}$ and $\sigma_{\nu}$ is simply
\begin{equation}
f_{\mu\nu}({\bf r})=-\Theta(\sigma_{\mu\nu}-|x|)
  \Theta(\sigma_{\mu\nu}-|y|)\Theta(\sigma_{\mu\nu}-|z|),
\label{Mayer}
\end{equation}
$\Theta(u)$ being the Heaviside step function ($=1$ for $u\geq 0$ and
$=0$ otherwise), and $\sigma_{\mu\nu}=(\sigma_{\mu}+\sigma_{\nu})/2$.
Therefore, since the Fourier transform of any of the step functions
appearing in (\ref{Mayer}) is
\[
\int_{-\infty}^{\infty}dx\,e^{ikx}\Theta(\sigma_{\mu\nu}-|x|) =
\frac{1}{2}\left(\hat\tau_{\mu}(k)\hat\zeta_{\nu}(k)+
\hat\tau_{\nu}(k)\hat\zeta_{\mu}(k)\right),
\]
$\hat\tau_{\mu}(k)\equiv 2k^{-1}\sin(\sigma_{\mu}k/2)$ and
$\hat\zeta_{\mu}(k)\equiv 2\cos(\sigma_{\mu}k/2)$, then the Fourier
transform of the Mayer function can be expressed as
\begin{eqnarray}
\hat f_{\mu\nu}({\bf k}) & = &
\frac{1}{8}\left(\hat w^{(3)}_{\mu}({\bf k})\hat w^{(0)}_{\nu}({\bf k})
+\hat w^{(0)}_{\mu}({\bf k})\hat w^{(3)}_{\nu}({\bf k})\right.
\nonumber \\
 & & \left.+\hat{\bf w}^{(2)}_{\mu}({\bf k})\cdot
\hat{\bf w}^{(1)}_{\nu}({\bf k})+\hat{\bf w}^{(1)}_{\mu}({\bf k})\cdot
\hat{\bf w}^{(2)}_{\nu}({\bf k})\right),
\label{FMT-decomp}
\end{eqnarray}
where
\begin{mathletters}
\begin{eqnarray}
\hat w^{(3)}_{\mu}({\bf k}) & \equiv &
   \hat\tau_{\mu}(k_x)\hat\tau_{\mu}(k_y)\hat\tau_{\mu}(k_z),
   \label{w3} \\
\hat{\bf w}^{(2)}_{\mu}({\bf k}) & \equiv & \left(
   \hat\zeta_{\mu}(k_x)\hat\tau_{\mu}(k_y)\hat\tau_{\mu}(k_z),
   \hat\tau_{\mu}(k_x)\hat\zeta_{\mu}(k_y)\hat\tau_{\mu}(k_z),
   \right. \nonumber \\
   & & \left.
   \hat\tau_{\mu}(k_x)\hat\tau_{\mu}(k_y)\hat\zeta_{\mu}(k_z)
   \right), \label{w2} \\
\hat{\bf w}^{(1)}_{\mu}({\bf k}) & \equiv & \left(
   \hat\tau_{\mu}(k_x)\hat\zeta_{\mu}(k_y)\hat\zeta_{\mu}(k_z),
   \hat\zeta_{\mu}(k_x)\hat\tau_{\mu}(k_y)\hat\zeta_{\mu}(k_z),
   \right. \nonumber \\
   & & \left.
   \hat\zeta_{\mu}(k_x)\hat\zeta_{\mu}(k_y)\hat\tau_{\mu}(k_z)
   \right), \label{w1} \\
\hat w^{(0)}_{\mu}({\bf k}) & \equiv &
   \hat\zeta_{\mu}(k_x)\hat\zeta_{\mu}(k_y)\hat\zeta_{\mu}(k_z).
   \label{w0}
\end{eqnarray}
\label{ws}
\end{mathletters}

Now that we have identified the so-called fundamental measures, the
FMT postulates the following excess (over the ideal one) free-energy
functional of the set of density profiles for each component
\cite{rosenfeld1}:
\begin{equation}
\beta F^{\rm ex}\left[\{\rho_{\mu}({\bf r})\}\right]=\int d{\bf r}\,
\Phi\left(\{n_i({\bf r}),{\bf n}_j({\bf r})\}\right)
\label{Fex}
\end{equation}
($\beta^{-1}=k_BT$) where $\Phi$ is assumed to be a function of the
following weighted densities:
\begin{mathletters}
\begin{eqnarray}
n_0({\bf r}) & \equiv & \sum_{\mu}\int\rho_{\mu}({\bf r}')
    \frac{1}{8}w^{(0)}_{\mu}({\bf r}-{\bf r}')\,d{\bf r}',
    \label{n0}  \\
{\bf n}_1({\bf r}) & \equiv & \sum_{\mu}\int\rho_{\mu}({\bf r}')
    \frac{1}{24}{\bf w}^{(1)}_{\mu}({\bf r}-{\bf r}')\,d{\bf r}',
    \label{n1}  \\
{\bf n}_2({\bf r}) & \equiv & \sum_{\mu}\int\rho_{\mu}({\bf r}')
    {\bf w}^{(2)}_{\mu}({\bf r}-{\bf r}')\,d{\bf r}'
    \label{n2} \\
n_3({\bf r}) & \equiv & \sum_{\mu}\int\rho_{\mu}({\bf r}')
    w^{(3)}_{\mu}({\bf r}-{\bf r}')\,d{\bf r}',
    \label{n3}
\end{eqnarray}
\label{ns}
\end{mathletters}
(the numerical coefficients have been chosen for convenience). The
uniform limit of $n_i({\bf r})$ is $\xi_i$ ($i=0,3$), and that of
${\bf n}_j({\bf r})$ is $\xi_j{\bf u}/3$ ($j=1,2$), where
${\bf u}\equiv(1,1,1)$ and the $\xi$s are the average density, mean
radius, surface and volume of the cubes:
\begin{equation}
(\xi_0,\xi_1,\xi_2,\xi_3)=\sum_{\mu}\left(1,\frac{\sigma_{\mu}}{2},
6\sigma_{\mu}^2,\sigma_{\mu}^3\right)\rho_{\mu}.
\label{xis}
\end{equation}
The $i$th averaged density ($i=0,1,2,3$) is a function with dimensions
(volume)$^{(i-3)/3}$; therefore $\Phi$, whose dimensions are
(volume)$^{-1}$, can be expanded as
\begin{equation}
\Phi=an_0+{\sf A}{\bf n}_1{\bf n}_2+{\sf B}{\bf n}_2{\bf n}_2
{\bf n}_2,
\label{expan}
\end{equation}
where $a$, ${\sf A}$ and ${\sf B}$ are, respectively, a scalar, a
2-index tensor and a 3-index tensor, all dependent on $n_3$. The
scaled-particle prescription imposes on $\Phi$ the following
differential equation \cite{rosenfeld1}:
\begin{equation}
-\Phi+\sum_{\mu}\rho_{\mu}\frac{\partial\Phi}{\partial\rho_{\mu}}+
n_0=\frac{\partial\Phi}{\partial n_3},
\label{differen}
\end{equation}
which yields $a=a_0-\ln(1-n_3)$, ${\sf A}=(1-n_3)^{-1}{\sf A}_0$
and ${\sf B}=(1-n_3)^{-2}{\sf B}_0$,
with $a_0$, ${\sf A}_0$ and ${\sf B}_0$ denoting a constant scalar,
a constant 2-index tensor and a constant 3-index tensor. The vanishing
of the excess free-energy per particle in the zero density limit leads
to $a_0=0$. On the other hand, the uniform limit of this free-energy
imposses a simplified form for the tensors:
\begin{mathletters}
\begin{eqnarray}
({\sf A}_0)_{ij} & = & \alpha_1{\bf u}_i{\bf u}_j
+\gamma_1\delta_{ij},  \\
({\sf B}_0)_{ijk} & = & \alpha_2{\bf u}_i{\bf u}_j{\bf u}_k
+\gamma_2{\bf u}_i\delta_{ij},
\end{eqnarray}
\end{mathletters}
$\alpha_{1,2}$ and $\gamma_{1,2}$ being numerical coefficients. Then
\begin{eqnarray}
\Phi & = & -n_0\ln(1-n_3)+\frac{\alpha_1n_1n_2+\gamma_1{\bf n}_1\cdot
{\bf n}_2}{1-n_3}  \nonumber \\
 & & +\frac{\alpha_2n_2^3+\gamma_2n_2{\bf n}_2\cdot{\bf n}_2}
{(1-n_3)^2},
\label{phi2}
\end{eqnarray}
where $n_i\equiv{\bf n}_i\cdot{\bf u}$ ($i=1,2$).

Now we have to routes to obtain the DCF of the fluid mixture: (i) from
Eqs.\ (\ref{Fex}), (\ref{ns}) and (\ref{phi2}), as \cite{evans}
\begin{equation}
c_{\mu\nu}({\bf r}-{\bf r}')=-\left.\beta
\frac{\delta F^{\rm ex}\left[\{\rho_{\mu}({\bf r})\}\right]}
{\delta\rho_{\mu}({\bf r})\delta\rho_{\nu}({\bf r}')}
\right|_{\{\rho_{\mu}({\bf r})\}=\{\rho_{\mu}\}},
\label{DCF}
\end{equation}
and (ii) from Eq.\ (\ref{FMT-DCF}), by explicitly obtaining the
involved functions. This latter expression can easily be computed
because the overlap region of two parallel cubes of sides
$\sigma_{\mu}$ and $\sigma_{\nu}$ is a parallelepiped of sides
$L_{\mu\nu}(x)$, $L_{\mu\nu}(y)$ and $L_{\mu\nu}(z)$ (provided they
do overlap, i.e.\ $f_{\mu\nu}({\bf r})=-1$), where
\begin{equation}
L_{\mu\nu}(s)=\left\{\begin{array}{l@{\quad\mbox{ if }}l}
\sigma_{\mu\nu}-\lambda_{\mu\nu} & |s|\leq\lambda_{\mu\nu} \\
\sigma_{\mu\nu}-|s| & |s|>\lambda_{\mu\nu}
\end{array}\right.
\label{Lmunu}
\end{equation}
($\lambda_{\mu\nu}\equiv|\sigma_{\mu}-\sigma_{\nu}|/2$). Therefore
\begin{mathletters}
\begin{eqnarray}
V_{\mu\nu}({\bf r}) & = & L_{\mu\nu}(x)L_{\mu\nu}(y)L_{\mu\nu}(z),
\label{V} \\
S_{\mu\nu}({\bf r}) & = & 2\{ L_{\mu\nu}(x)L_{\mu\nu}(y)+
   L_{\mu\nu}(x)L_{\mu\nu}(z)  \nonumber \\
           & & +L_{\mu\nu}(y)L_{\mu\nu}(z) \},
\label{S}  \\
R_{\mu\nu}({\bf r}) & = & \frac{1}{6}\left\{ L_{\mu\nu}(x)+
                 L_{\mu\nu}(y)+L_{\mu\nu}(z) \right\}.
\label{R}
\end{eqnarray}
\label{VSR}
\end{mathletters}
By matching both expressions of the DCF it follows that $\alpha_1=0$,
$\gamma_1=3$, $\alpha_2=5/432$ and $\gamma_2=-1/48$. Then
\begin{mathletters}
\begin{eqnarray}
\chi_0 & = & \frac{1}{1-\xi_3},   \label{chi0} \\
\chi_1 & = & \frac{\xi_2}{(1-\xi_3)^2},   \label{chi1} \\
\chi_2 & = & \frac{\xi_1}{(1-\xi_3)^2}+\frac{1}{36}
\frac{\xi_2^2}{(1-\xi_3)^3},  \label{chi2} \\
\chi_3 & = & \frac{\xi_0}{(1-\xi_3)^2}+\frac{2\xi_1\xi_2}{(1-\xi_3)^3}+
\frac{1}{36}\frac{\xi_2^3}{(1-\xi_3)^4},  \label{chi3}
\end{eqnarray}
\label{chis}
\end{mathletters}
and the excess free-energy per unit volume turns out to be, in the
uniform limit,
\begin{equation}
\Phi=-\xi_0\ln(1-\xi_3)+
\frac{\xi_1\xi_2}{1-\xi_3}+\frac{1}{216}\frac{\xi_2^3}{(1-\xi_3)^2}.
\label{phiunif}
\end{equation}
The compressibility equation of state
\begin{equation}
\beta\frac{\partial P^{c}}{\partial\rho_{\mu}}=1-\sum_{\mu}\rho_{\mu}
\hat c_{\mu\nu}({\bf 0})
\label{compress}
\end{equation}
leads to
\begin{equation}
\beta P^{c}=\frac{\xi_0}{1-\xi_3}+\frac{\xi_1\xi_2}{(1-\xi_3)^2}+
\frac{1}{108}\frac{\xi_2^3}{(1-\xi_3)^3}.
\label{Pc}
\end{equation}
On the other hand, the virial theorem provides a different expression
for the pressure:
\begin{equation}
\beta P^{v}=\xi_0+\frac{1}{6}\sum_{\mu,\nu}\rho_{\mu}\rho_{\nu}
\int d{\bf r}\, {\bf r}\cdot\nabla f_{\mu\nu}({\bf r})
g_{\mu\nu}({\bf r}),
\label{virial}
\end{equation}
with $g_{\mu\nu}({\bf r})$ the pair correlation function.
Straightforward manipulations lead to
\begin{equation}
\beta P^{v} = \xi_0-4\sum_{\mu,\nu}\rho_{\mu}\rho_{\nu}
\sigma_{\mu\nu}
\int_0^{\sigma_{\mu\nu}}\!\!dy\int_0^{\sigma_{\mu\nu}}\!\!dz\,
c_{\mu\nu}(\sigma_{\mu\nu}^-,y,z),
\label{Pv}
\end{equation}
from which it follows that $P^v=P^c$. The expression found for the DCF
of this system has thus the remarkable property of being
{\em thermodynamically consistent} without having impossed such a
feauture anywhere in the approximate scheme. Furthermore, this (unique)
equation of state matches the second and third virial coefficients,
which can be readily computed to be $B_2\rho^2=\xi_0\xi_3+
\frac{2}{3}\xi_1\xi_2$ and $B_3\rho^3=\xi_0\xi_3^2+2\xi_1\xi_2\xi_3+
\xi_2^3/108$ (where $\rho=\sum_{\mu}\rho_{\mu}=\xi_0$), but this is
implicitly assumed in the theory \cite{rosenfeld1}.

A direct comparison with the simulations of Ref.\ \cite{dijkstra}
is meaningless, because the latter are performed on a lattice system
with a lattice spacing comparable to the small particle size; moreover,
there are no other results of this kind available
for this model, so the accuracy of the result just presented has
still to be tested. For the one-component system, however, both
the exact virial coefficients and the virial expansion of
approximate theories (such as PY) are known up to seventh order
\cite{7virial}. From them we can see
that the values obtained from Eq.\ (\ref{Pc}) neither match the
exact ones nor those obtained from the PY approximation, from the
fourth coefficient on. Thus we can conclude that the present results
are --- as expected --- approximate, but they differ from the PY
approximation, contrary to what happens for hard spheres (the
equivalence for hard spheres might indeed be an exception rather than
a rule).

But the most striking consequence which follows from the results
presented thus far is the prediction of a spinodal instability of a
binary mixture of hard cubes for large enough side ratio. Furthermore,
this spinodal can be analytically computed within this approximation.
Let us see how it can be done. The instability is caused by a
violation of the concavity of the free-energy. The concavity of this
function can be expressed as the possitive definiteness of the hessian
matrix of the free-energy per unit volume, $f\equiv F/V$, as a
function of the partial densities, $\rho_{\nu}$. Since $f$ is given by
$f=\sum_{\nu}\rho_{\nu}\mu_{\nu}-P$, with $\mu_{\nu}$ the chemical
potential of the $\nu$ component, by using
the Gibbs-Duhem relation we can obtain \cite{lebowitz}:
\begin{equation}
M_{\nu\lambda} \equiv
  \beta\frac{\partial^2f}{\partial\rho_{\nu}\partial\rho_{\lambda}} =
  \beta\frac{\partial\mu_{\nu}}{\partial\rho_{\lambda}} =
  \frac{1}{\rho_{\nu}}\delta_{\nu\lambda}-\hat c_{\nu\lambda}({\bf 0}).
\label{matrixM}
\end{equation}
From the expression for the DCF the $2\times 2$ matrix $M$ of
the binary mixture can be readily computed, and the
stability condition follows from the possitiveness of
its determinant --- since all its elements are possitive. By defining
$\eta_{\nu}\equiv\sigma_{\nu}^3\rho_{\nu}$,
the packing fraction of species $\nu$; $\eta\equiv\eta_1+\eta_2=\xi_3$,
the total packing fraction of the fluid; $r\equiv\sigma_1/\sigma_2$,
the large-to-small side ratio ($r\geq 1$); and $x\equiv\eta_1/\eta$,
the relative packing fraction of the large component, a tedious but
straightforward calculation yields
\begin{equation}
\rho_1\rho_2|M|=\frac{\eta^2}{(1-\eta)^4}\left[1+\frac{4}{\eta}
+\frac{1}{\eta^2}-\frac{3(r-1)^2}{r}x(1-x)\right],
\label{det}
\end{equation}
and thus the mixture is stable provided the expression within
square braquects is possitive.

An immediate consequence of this result is that the mixture is stable
if and only if $1\leq r<5(1+\sqrt{1-1/25})\approx 9.98$.
Above this threshold there is a region where the mixture phase
separates (see figure), limited below by the curve
\begin{equation}
\eta^{-1}=3^{1/2}\left[1+\frac{(r-1)^2}{r}x(1-x)\right]^{1/2}-2.
\label{spinodal}
\end{equation}

\begin{figure}
  \begin{center}
  \setlength{\unitlength}{1cm}
  \hspace*{-4cm}
    \begin{picture}(10,11.5)
      \epsfxsize=13cm
      \epsffile{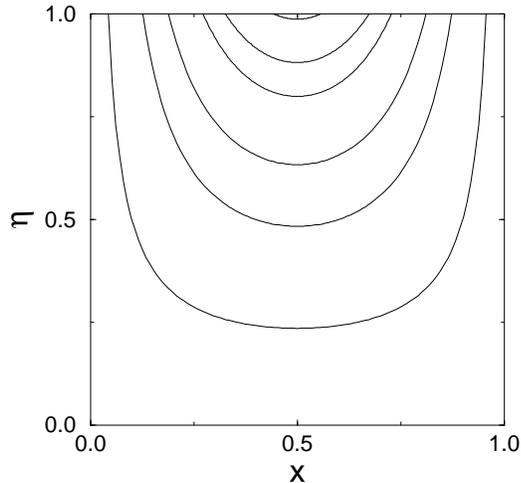}
    \end{picture}
  \end{center}
  \vspace*{-5cm}
\label{spinodalfig}
\caption[]{Demixing phase diagram for a binary mixture of parallel
hard cubes:
$\eta$ is the total packing fraction and $x$ the relative packing
fraction of the large cubes. Curves represent the spinodal lines for
side ratios 10, 11, 12, 15, 20 and 50 (from top to bottom).}
\end{figure}

Thus we have achieved a qualitative agreement with the simulations of
Ref.\ \cite{dijkstra}; however, a direct comparison of the results is,
perhaps, too na\"\i ve. In the simulations the authors find demixing
for $r\geq 3$, while this calculations shift this value
up to $\approx 10$. But the simulations are performed on a lattice,
and then the size of the small particle (2 lattice spacings in this
case) is also important, for a larger size with the same size ratio
implies a larger number of accessible sites on the lattice for
the small particles, with the corresponding gain in entropy.
Therefore, if the small particles are larger, it is plausible that the
stability of the mixture increases to larger size ratios. The results
for the continuous model we are dealing with in this letter would
correspond to the limit $\sigma_1\to\infty$ and $\sigma_2\to\infty$,
while keeping $\sigma_1/\sigma_2$ constant, what explains its larger
stability. Some simulations could ascertain this conclusion.

The two-dimensional case, i.e.\ a mixture of hard squares, is a simple
reproduction of the calculations here presented. This is something
unusual in view of what happens for hard spheres \cite{rosenfeld1},
for which the FMT can be applied only in odd dimensions. For cubes the
calculations can be done in a similar way no matter the dimension. In
two dimensions, for instance, the form of the DCF is that of Eq.\
(\ref{FMT-DCF}) except for the volume term, which is logically missing.
Again the equation of state reproduces the three first terms of the
virial expansion, as expected from the FMT. The novelty is that it can
be shown that the binary mixture of hard squares is {\em always}
stable, whichever the size ratio. It is interesting to note that the
authors of Ref.\ \cite{dijkstra} arrived to the same conclusion in
their simulations. Full details on the hard-square mixture will be
given elsewhere \cite{cuesta}.

In summary, I have provided in this letter the DCF of a new model,
namely a mixture of hard cubes, obtained by means of the FMT, an
approximation introduced
by Rosenfeld and thus far applied only to reproduce the results of a
mixture of hard spheres. Apart from the importance of having analytical
results for new models of liquids, the relevance of this calculations
relies on the fact that a spinodal instability is predicted for the
binary mixture, in qualitative agreement with the simulations performed
on an equivalent lattice model. The calculations may serve as a
reference for further studies on related system, for instance, the
orientational freezing of liquid crystals, if the model is extended to
a mixture of parallelepipeds, or the polymer collapse, of which Ref.\
\cite{dijkstra} reports some simulations. Of course it would also be
interesting to apply this results to the lattice system, in order to
see whether they can reproduce the simulations quantitatively.

\end{multicols}

\end{document}